# Improving Android Malware Detection Through Data Augmentation Using Wasserstein Generative Adversarial Networks

Kawana Stalin

Department of Software Engineering

Northwestern Polytechnical University

China

stalinkawana@mail.nwpu.edu.cn

Mikias Berhanu Mekoya

Department of Software Engineering

Northwestern Polytechnical University

China

***Abstract*** -- Generative Adversarial Networks (GANs) have demonstrated their versatility across various applications, including data augmentation and malware detection. This research explores the effectiveness of utilizing GAN-generated data to train a model for the detection of Android malware. Given the considerable storage requirements of Android applications, the study proposes a method to synthetically represent data using GANs, thereby reducing storage demands. The proposed methodology involves creating image representations of features extracted from an existing dataset. A GAN model is then employed to generate a more extensive dataset consisting of realistic synthetic grayscale images. Subsequently, this synthetic dataset is utilized to train a Convolutional Neural Network (CNN) designed to identify previously unseen Android malware applications. The study includes a comparative analysis of the CNN's performance when trained on real images versus synthetic images generated by the GAN. Furthermore, the research explores variations in performance between the Wasserstein Generative Adversarial Network (WGAN) and the Deep Convolutional Generative Adversarial Network (DCGAN). The investigation extends to studying the impact of image size and malware obfuscation on the classification model's effectiveness. The data augmentation approach implemented in this study resulted in a notable performance enhancement of the classification model, ranging from 1.5% to 7%, depending on the dataset. The highest achieved F1 score reached 0.975.

***Keywords***--Generative Adversarial Networks, Android Malware, Data Augmentation, Wasserstein Generative Adversarial Network

## 1. Introduction

Android is the most popular operating system for smartphones. The Android operating system offers a wide range of free features that are open source which has made it widely preferred among users [1]. The constantly growing Android user base, coupled with the vast array of device models and configurations, presents a challenging platform for security personnel. A notable component of the Android system is that applications are accessible not only from the local Google Play Store but also through an assortment of third-party application sources [2]. This open nature has fostered innovation and a vibrant app ecosystem but also makes it susceptible to vulnerability exploitation. Android malware refers to malicious software specifically designed to compromise the security and functionality of Android devices. These malicious programs come in various forms, ranging from obfuscated and sophisticated attacks to more overt and disruptive tactics. The motivations behind Android malware are diverse, encompassing financial gain, sensitive data theft, espionage, and even political objectives. In 2018, apps that could put users, user data, or devices at risk, also known as Potentially Harmful Applications (PHAs) comprised 0.11% of app installations downloaded outside of Google Play [3]. As Android malware continues to evolve in sophistication and adaptability, it highlights the critical need for robust security measures, and collaboration between device manufacturers, app developers, security researchers, and users to mitigate the risks and safeguard the Android ecosystem.

In the realm of cybersecurity, where the landscape of malicious software constantly evolves, the availability of diverse and representative datasets is crucial for training accurate and resilient models. Data augmentation has become an important topic in the field of machine learning as it plays a pivotal role in enhancing the robustness and effectiveness of malware detection models. It enables model accuracy to be improved with limited data. Traditional data augmentation techniques involve applying various transformations to the existing malware samples, such as rotations, flips, or changes in scale, thereby diversifying the dataset without altering the underlying malicious characteristics. Recently, two of the most popular and promising new data augmentation techniques are the Generative Adversarial Network (GAN) and Variational Autoencoder (VAE) [4] models, which have been employed in some scientific studies to virtually expand the sample size of clinical investigations, thereby mitigating costs, time constraints, dropout rates, and ethical concerns [5]. GANs are the latest unsupervised deep-learning approach that can be used to create newly synthesized instances of data in the form of images, music, video, speech, text, etc. The learning process is unsupervised in the sense that the networks do not rely on labeled training data, and there is no explicit guidance provided for the generator to generate specific outputs. The original idea of GANs was first introduced by Ian Goodfellow et al., in 2014 [6]. We find that GANs are a suitable selection for our approach, particularly due to their effectiveness with image data, allowing us to customize and integrate them seamlessly with the architecture of a Convolutional Neural Network (CNN). A typical GAN's deep neural network consists of two components: a generator and a discriminator. The generator creates realistic data from a latent space and the discriminator determines whether the generated data is genuine or not. During training, the generator progressively becomes better at creating images that look real, while the discriminator becomes better at telling them apart. The process reaches equilibrium when the discriminator can no longer distinguish real images from fakes. If the discriminator is unable to distinguish between real and fake data, then it is assumed that the generated data is real. Figure 1 shows the basic principle of GANs.

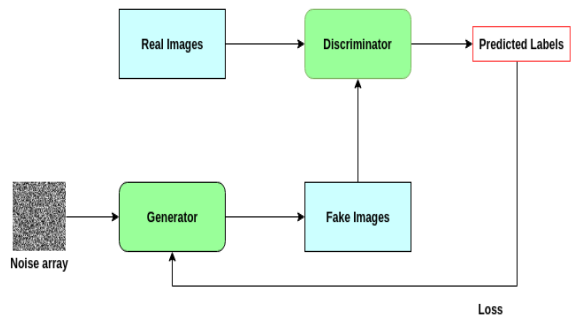

Figure 1. GAN Concept

By using GANs, researchers can create new datasets from existing malware samples, allowing them to train machine learning models on larger, more diverse datasets and improve the accuracy of their models, thereby giving security professionals the leverage to better identify and respond to potential threats. This powerful machine-learning tool has been used for a variety of tasks, including the detection of Android malware. GANs have been successfully applied in several Android malware detection systems, demonstrating their utility for this task. GAN-based models have shown promising performance in detecting Android malware, outperforming traditional machine-learning approaches in certain cases.

However, Generative Adversarial Networks are challenging to train. They require thorough regularization, extensive computational resources, and expensive hyper-parameter sweeps [7]. We, therefore, need a mechanism to accurately evaluate their performance. The research gap in this area is the lack of robust evaluation metrics for GAN-based malware analysis methods. Currently, there is no standard way to measure the performance of these methods, which makes it difficult to compare the results of different approaches and determine which ones are most effective. Another research gap is ways of improving the quality and diversity of the synthetic malware representations generated by the generator network, as well as developing better strategies for training the discriminator network to accurately distinguish between synthetic and real malware.

Malware can be difficult to detect and analyze. One approach to malware analysis is by representing malware samples as images. Image-based representation of malware has shown promising results in previous research. This approach has gained popularity in recent years due to its ability to leverage the power of computer vision and image-processing techniques to analyze malware. Traditional machine-learning classifiers can be misled by carefully injecting crafted data into a training set [8]. This challenge can be addressed by employing image-based detection models, as image-processing techniques are intentionally designed to withstand noise and various types of distortion. This inherent robustness enhances their resilience against some obfuscation techniques [9] employed by malware authors to evade detection. Image representation of malware usually involves converting malware samples into grayscale [10], or color images, where each pixel in the image represents a specific feature of the malware. These features can include opcode sequences, file header information, and other attributes that are characteristic of the malware. Once converted to images, the malware samples can be analyzed using various computer vision and image processing techniques, such as convolutional neural networks and image segmentation algorithms [11].

The magnitude of certain Android installation files can reach up to 500 megabytes, posing a challenge for researchers who must download thousands of apps for training purposes. This undertaking not only demands considerable time but also necessitates extensive storage space. For instance, acquiring over 40,000 apps for the present study spanned a duration of approximately three weeks and consumed more than 100 gigabytes of storage. To mitigate these resource-intensive requirements, employing images as representations of malware proves advantageous due to their reduced storage footprint. Thousands of synthetic image representations of malware samples can be generated through a Generative Adversarial Network (GAN), with the initial input of only 1000 or fewer real samples, a potential reduction in both time and storage usage can be achieved. The primary inquiry in this study revolves around the impact and effectiveness of such generated data on the performance of the classification model, addressing pertinent questions in this domain.

The rest of this paper is organized as follows: Section 2 explains the background and related work. Section 3 explains our approach. Section 4 presents the datasets. Section 5 presents the experimental results. Finally, we conclude in Sections 6 and 7.

## 2. Related Work

**2.1 Image-Based Representation of Malware**
To the best of our knowledge, the earliest research on the image-based representation of malware was in 2011 by L. Nataraj et al [12]. The authors employed grayscale images to represent malware binaries and noted that, for numerous malware families, images associated with the same family exhibit striking similarities in both layout and texture. In recent years, there has been a notable recognition of the undeniable potential of using image-based techniques to represent malware. In [13], Mazhar Javed Awan et al. proposed a spatial attention and convolutional neural network based on a deep learning framework for image-based classification of 25 well-known malware families. Their proposed model achieved an F1 score higher than 97%. Duc-Ly Vu et al. [14], developed a novel hybrid image transformation method to convert malware binaries into color images that convey binary semantics. A deep network trained on these features achieved a performance of 99.14% in terms of accuracy. N. Daoudi et al. [15], proposed DexRay, which converts the bytecode of the app DEX (Dalvik Executable) files into grey-scale "vector" images and feeds them to a 1-dimensional Convolutional Neural Network model. While simple, their approach is effective with an F1 score of 0.96. However, this approach did not perform well in the obfuscated applications detection when the training set does not include obfuscated applications. Yuxin Ding et al [16], proposed a detection model that analyzes malware features and designs the feature representation of malware by extracting a bytecode file from an Android APK file and converting it into a two-dimensional bytecode matrix, then using a Convolution Neural Network (CNN), to train a detection model to classify malware. The authors claim their approach is effective, especially at detecting malware encrypted using polymorphic techniques. F.M. Darus et al [17], extracted Classes.dex files from the Android APK files before they were converted into images. Utilizing three machine learning algorithms, their findings indicate that classification based on the data section outperforms classification on entire classes.dex files.

**2.2 Data Augmentation Using Generative Adversarial Networks**

From 2017 onward, there has been a notable increase in the number of studies solely employing GANs, with approximately 50% of these studies concentrating on image synthesis. Among these applications, cross-modality image synthesis stands out as the most significant [18]. Some recent studies have also explored the use of GANs for data augmentation for Android malware detection. In a comparative review by Roland Burks III et al. [19], the authors claimed that adding synthetic malware samples generated by GAN to the training data improved the accuracy of the Residual Network (ResNet-18) classifier by 6%, compared to 2% by VAE. In a study by Chen et al. [20], GANs were used to augment malware samples from the DroidKungFu and Geinimi families. The authors tested four different architectures (DCGAN, WGAN, CGAN, and CycleGAN) on their dataset, with CycleGAN producing the best accuracy. They also found that combining synthetic samples with real-world samples improved the F1 score by 5 ~ 20%. In another study by Wang et al. [21], the authors proposed a general framework named AdvAndMal, which consists of a two-layer network for adversarial training to generate adversarial samples and improve the effectiveness of the classifiers in Android malware detection and family classification. Their experimental results on 12 families with the largest number of samples in the Drebin dataset showed an increase in the overall accuracy of the framework of 1 ~ 2%. Overall, recent studies have demonstrated that data augmentation using GANs can be effective in improving model accuracy on Android malware detection tasks. This could potentially help reduce false positives and false negatives when detecting malicious applications on mobile devices. Additionally, these studies suggest that combining both real-world and synthetic samples can further improve model performance compared with just using synthetic samples alone. However, this approach to malware detection and classification is never without challenges. So far, we have identified the limitations associated with the use of GANs in the context of Android malware detection as follows;

1. The opcode sequence length extracted from each program sample varies. In cases where the opcode count is either too low or too high, the eventual uniform normalization process may result in image distortion due to excessive adjustments, consequently influencing the efficacy of model learning [20].

2. Deep learning applications utilize feed-forward neural network structures that require a fixed-size input and map it to a fixed-size output [22].

3. Pre-existing deep learning architectures for images require a standard image size.

4. GANs require a large amount of training data in order to generate accurate synthetic samples, which may be difficult or expensive to obtain in some cases.

5. GANs necessitate model variation for acquiring data distribution via unsupervised learning and generating authentically realistic synthetic samples [23]. This process demands substantial computational resources and time.

6. The lack of easy-to-implement metrics to evaluate GAN performance.

7. Other challenges include instability during training and a lack of interpretability of the generated samples, which can make it difficult to identify potential vulnerabilities or false positives in the generated data.

While there are some challenges associated with this technique, it appears to be a promising approach for improving model accuracy and detecting unknown threats in mobile environments.

The following are the contributions of this study:

1. We overcome limitations 1 and 2 by using six different image sizes and comparing them to see which size produces the best results.
2. We investigate the feasibility of using only synthetic images as well as mixing synthetic images with real ones to train the classification model and test it on the real data.
3. We generate image representation of malware using two different GAN models (WGAN & DCGAN) and compare the performance of the classification model trained on images generated by the two.
4. We overcome limitation 4 by using a custom loss function to evaluate GAN performance and the FID metric to ensure only good-quality images are saved at each iteration.

## 3. Approach

### 3.1 Image Representation of Android Malware

A Python library called Androguard [24] is used as the main tool to interact with Android Files. Static analysis of Android apps can be done by extracting the bytecode i.e. the Classes.dex (Dalvik Executable), and AndroidManifest.xml from APK (Android Package) installation files. The contents of these files are converted into grey-scale vector images. The Classes.dex file in the APK is the compressed file which is made up of all the Java classes and native libraries in the application code [25] [26]. These contain all the operating instructions of the application and runtime data. AndroidManifest.xml contains a lot of features that can be used for both static and dynamic analysis which include [27]:

- Activities - An Android activity refers to a singular screen within the user interface of an Android application.
- Broadcast receivers and providers
- Metadata – This represents an extra option for storing information that is accessible across the entire application.
- The permissions requested by the application.
- System features such as camera and internet.

AndroidManifest.xml files are an important asset in Android malware analysis because authors of Android malicious applications often use software packers to protect themselves against decompilation. In such a case, even if we successfully decompile the application, the only available file is usually the manifest file [28].

We focus on static analysis because it can be performed by simply extracting the bytecode instead of running the application in a real environment. Dynamic analysis requires analyzing the application while it is running in a simulated or real environment which costs time and resources. To do static analysis, we first store image representations of both the Classes.dex and AndroidManifest.xml files from the unobfuscated applications. To convert Classes.dex file into images, we leverage the algorithm proposed by Nadia Daoudi et al [15] as shown in Figure 2.

**Algorithm 1**: Algorithm describing 8-bit grey-scale "vector" image generation from an APK
**Input:** APK file
**Output:** 8-bit grey-scale "vector" image of size (1, 128x128)

bytestream ← ∅
for *dexFile in APK* do
| bytestream ← bytestream + dexFile.toByteStream()
end
l ← bytestream.length()
img ← generate8bitGreyScaleVectorImage(bytestream, l)
img.resize_to_size(height=1, width=128x128)
img.save()

Figure 2. Algorithm for generating a grey-scale image from an APK

To generate images of a different width, all we do is make minor changes to lines 2 and 9 of Algorithm 1 in Figure 2 accordingly. We then repeat the same process for obfuscated applications. Figure 3 shows images generated from Classes.dex and AndroidManifest.xml files. The images depicting malware and benign instances for Classes.dex files exhibit clear visual distinctions discernible to the naked eye, whereas the images associated with AndroidManifest.xml files lack such obvious differences as shown below. How this affects the performance of the classification model and the reasons why are discussed in section 5.3.

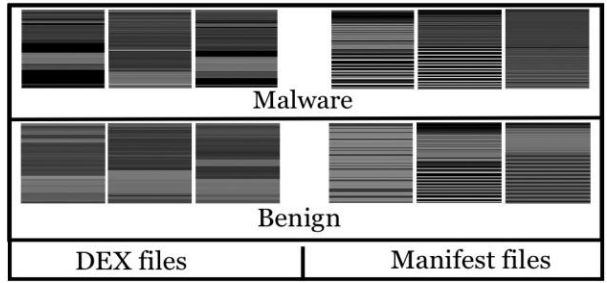

Figure 3. Examples of generated images: Malware vs. benign

### 3.2 Obfuscation Process

In simple words, obfuscation is the action of making something obscure, unclear, or unintelligible. Hackers may deliberately obfuscate code to conceal its purpose, in order to prevent reverse engineering or detection. In our case, obfuscation was achieved using the state-of-the-art Android application obfuscator Obfuscapk, which is a modular Python tool for obfuscating Android applications without needing their source code [29]. To ensure that the resulting application is highly obfuscated, we employed eleven different obfuscation processes:

1. AdvancedReflection - Invokes dangerous APIs of the Android Framework using reflection.
2. ArithmeticBranch - Inserts junk code.
3. AssetEncryption - Encrypts asset files.
4. CallIndirection - Modifies the control-flow graph without impacting the code semantics.
5. ClassRename - Changes the package name and renames classes.
6. ConstStringEncryption - Encrypt constant strings in code.
7. FieldRename - Renames fields.
8. LibEncryption - Encrypts native libs.
9. MethodOverload - Exploits the overloading feature of the Java programming language to assign the same name to different methods but using different arguments.
10. MethodRename - Renames methods.
11. Reorder - Changes the order of basic blocks in the code.

Due to unknown errors, not all applications were successfully obfuscated using all of the above obfuscation processes. For the applications that underwent successful obfuscation, we ensured the implementation of a minimum of five obfuscation techniques from the aforementioned list for each application. About 50% of the applications from each dataset are selected for obfuscation and mixed back into the original dataset.

### 3.3 Data Augmentation

As stated in [20], most image data augmentation research in recent years has focused on the Deep Convolutional GAN (DCGAN) due to its simplicity. In contrast, we utilize the Wasserstein GAN (WGAN) and compare it against the DCGAN. WGAN is an alternative to traditional GAN training first introduced by Martin Arjovsky, et al. [30], in 2017. A WGAN can easily be implemented by taking the standard DCGAN and applying a few minor changes. For example, instead of using a discriminator to classify or predict the probability of generated images as being real or fake, the WGAN changes or replaces the discriminator model with a critic that scores the realness or fakeness of a given image. This change is motivated by a theoretical argument that training the generator should seek a minimization of the distance between the distribution of the data observed in the training dataset and the distribution observed in generated examples [31]. Although it sometimes takes slightly longer than the standard DCGAN to train, the benefit of the WGAN is that the training process is more stable and less sensitive to model architecture and choice of hyperparameter configurations. However, due to the use of weight clipping to enforce a Lipschitz constraint on the critic, the WGAN sometimes fails to converge or generates samples of inferior quality. Ishaan Gulrajani, et al. [32], proposed a way of overcoming this disadvantage i.e. penalizing the norm of the gradient of the critic with respect to its input. Figure 4 shows the algorithm for the improved training of WGANs, as in [32].

**Algorithm 2** WGAN with gradient penalty. We use default values of $\lambda = 10$, $n_{\text{critic}} = 5$, $\alpha = 0.0001$, $\beta_1 = 0$, $\beta_2 = 0.9$.

**Require:** The gradient penalty coefficient $\lambda$, the number of critic iterations per generator iteration $n_{\text{critic}}$, the batch size $m$, Adam hyperparameters $\alpha, \beta_1, \beta_2$.
**Require:** initial critic parameters $w_0$, initial generator parameters $\theta_0$.

1: **while** $\theta$ has not converged **do**
2:   **for** $t = 1, ..., n_{\text{critic}}$ **do**
3:     **for** $i = 1, ..., m$ **do**
4:       Sample real data $x \sim \mathbb{P}_r$, latent variable $z \sim p(z)$, a random number $\epsilon \sim U[0,1]$.
5:       $\tilde{x} \leftarrow G_\theta(z)$
6:       $\hat{x} \leftarrow \epsilon x + (1 - \epsilon)\tilde{x}$
7:       $L^{(i)} \leftarrow D_w(\tilde{x}) - D_w(x) + \lambda(\|\nabla_{\hat{x}} D_w(\hat{x})\|_2 - 1)^2$
8:     **end for**
9:     $w \leftarrow \text{Adam}(\nabla_w \frac{1}{m} \sum_{i=1}^{m} L^{(i)}, w, \alpha, \beta_1, \beta_2)$
10:   **end for**
11:   Sample a batch of latent variables $\{z^{(i)}\}_{i=1}^{m} \sim p(z)$.
12:   $\theta \leftarrow \text{Adam}(\nabla_\theta \frac{1}{m} \sum_{i=1}^{m} -D_w(G_\theta(z)), \theta, \alpha, \beta_1, \beta_2)$
13: **end while**

Figure 4. Algorithm for Improved Training of WGANs

Previous research on data augmentation for Android malware detection didn't use any mechanism to accurately evaluate the performance of their GAN models or the quality of the generated images. In contrast, in our experiments, synthetic image representations of malware are generated using both the DCGAN and WGAN and their quality is measured using the Frechet Inception Distance (FID) at every iteration. The FID evaluates the quality of images generated by generative adversarial networks and how similar they are to the real ones better than the Inception Score (IS). It was first introduced by M. Heusel, et al. [33], in 2017. The lower the FID, the higher the quality of the images. However, FID can be biased depending on the model. Min Jin Chong et al. [34], showed how to extrapolate the score to obtain an effectively bias-free estimate (termed FID∞) of scores computed with an infinite number of samples, which requires good estimates of scores with a finite number of samples. Eyal Betzalel et al. [35], found that among inception-based metrics, FID∞ has the highest correlation with Kullback–Leibler (KL) and Reverse KL (RKL) both of which are a commonly used measure of the difference between probability distributions, indicating that it is a more reliable metric. Based on visual inspection of our generated datasets, enough image clarity is reached at FID∞ scores of 90 or less. Therefore, we ask the generator to continue generating new images until an FID∞ score of 90 or less is reached for each of the generated images, before finally saving them. We also evaluate the performance of the WGAN model by implementing a custom Wasserstein loss function using Keras that calculates the average score for real or fake images and plots it against the number of epochs. Results are discussed in section 5.1.

### 3.4 Deep Learning Architecture

Convolutional Neural Networks (CNNs) are a class of deep neural networks that have proven highly effective in tasks related to processing structured grid data, such as images [36]. Although we employ the same technique as in [15] to generate image representations of malware, we use a different approach in our deep learning architecture. Instead of 1-dimensional convolution layers, we employ 2-dimensional convolutional layers because image data is inherently 2-dimensional. 2-dimensional convolutional layers are specifically designed to capture spatial relationships and patterns in two-dimensional data, making them well-suited for image processing tasks. Using 2-dimensional convolutional layers allows the neural network to effectively learn hierarchical features in both horizontal and vertical dimensions, making it more appropriate for image-related tasks such as image classification, object detection, facial recognition, and more.

**3.5 Model Overview**

Approximately 25-30% of applications from each dataset (including obfuscated and unobfuscated) are randomly selected and used for adversarial training in the GAN model to generate more images to train the classification model. Three models were used for classification: Model 1, Model 2, and Model 3. Model 1 is a 2-dimensional CNN trained on real images. Real images also known as real data are the images that are generated directly from the application's bytecode, while synthetic images also known as GAN data are the images that are generated by the GAN model. Model 2 is also a 2-dimensional CNN, but it's trained on the dataset generated by the GAN model only (GAN data). Model 3 is trained on both real and synthetic images. All three models are tested using the remaining 70-75% of the original obfuscated and unobfuscated datasets, and compared against each other. The concepts of Models 1 and 2 are shown in Figure 5 and 6 respectively.

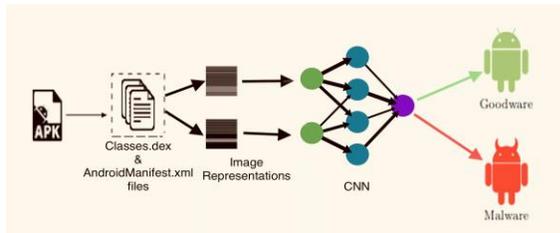 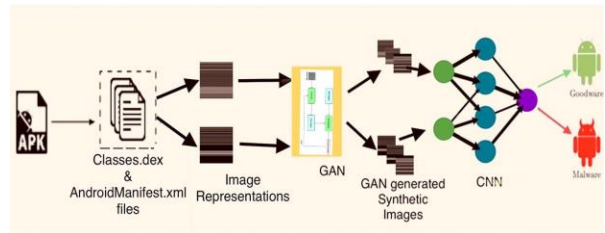

Figure 5. Model 1              Figure 6. Model 2

As seen in the figures above, Model 2 is a little more complex and takes more time than Model 1 as it requires one extra step in between before classification. This ensures that the data generated is more diverse, and hopefully, if we combine Models 1 and 2, the resulting model (Model 3) will be more capable of detecting never seen before malware and resilient to obfuscation. The results of this approach are discussed in Chapter 5.

## 4. Datasets

Most of the datasets used in the recent Android malware research community are outdated. In contrast, we used four datasets that are the latest, reliable, large-scale, and reflect current malware evolution trends. We also chose these datasets based on availability. The following are the four datasets:

**4.1 CICMalDroid 2020**

The first dataset we used is called CICMalDroid 2020 [37] [38] provided by the Canadian Institute for Cybersecurity. They collected more than 17,341 Android samples from several sources including VirusTotal service, Contagio security blog, AMD, MalDozer, and other sources mostly used in recent research. The samples were collected from December 2017 to December 2018. To verify the maliciousness, they scanned all the benign and malicious samples with VirusTotal. The properties of this dataset are summarized as follows [37] [38]:

   - **Big** - It has more than 17,341 Android samples.

   - **Recent** - It includes recent and sophisticated Android samples until 2018.

   - **Diverse** - It has samples spanning five distinct categories: Benign, Adware, Banking malware, SMS malware, and Riskware. All other applications that are not malicious are considered benign. Mobile Adware refers to the advertising material (i.e., ads) that typically hides inside legitimate apps that have been infected by malware. Adware can infect and root-infect a device, forcing it to download specific Adware types and allowing attackers to steal personal information. Mobile Banking malware is a specialized malware designed to gain access to the user's online banking accounts by mimicking the original banking applications or banking web interface. SMS malware exploits the SMS service as its medium of operation to intercept SMS payload for conducting attacks. Riskware refers to legitimate programs that can cause damage if malicious users exploit them.

- **Comprehensive** - It includes the most complete captured static and dynamic features compared with publicly available datasets.

This is the easiest to access of the four datasets.

### 4.2 Drebin

The Drebin dataset contains 5,560 Android applications from 179 different malware families. The samples have been collected from August 2010 to October 2012 [39] [40]. This dataset is the oldest and most out-of-date of the four, and the authors no longer maintain it. However, we need to include this particular dataset for comparison purposes with the recent research.

### 4.3 MalRadar

The MalRadar dataset is a growing and up-to-date Android malware dataset that contains 4,534 unique Android malware samples (including both apps and metadata) released from 2014 to April 2021. The authors of this dataset crawled all the mobile security-related reports released by ten leading security companies and used an automated approach to extract and label the useful ones describing new Android malware and containing Indicators of Compromise (IoC) information [41].

### 4.4 AndroZoo

AndroZoo [42], is a collection of Android applications collected from several sources, including the official Google Play app market. It currently contains 23,836,516 different applications at the time of writing. However, due to lack of time and storage, we could only download 8738 benign and 5670 malware applications from the AndroZoo repository with the dexdate starting from June 2019 and satisfy other specified criteria such as Virus Total rating, application size, and which markets they were downloaded from.

## 5. Experiments and Evaluation

Experiments in this study were implemented in Python programming language on a hardware configuration equipped with NVIDIA GeForce GTX 1650 graphics card and 16GB RAM.

### 5.1 How Much Do the Performances of The Two GAN Models Differ?

We conducted experiments to determine whether the WGAN and DCGAN can learn the characteristics of the real malware samples and generate additional synthetic samples. We compared the performance of the two GAN models as follows;

### 5.1.1 Visual Inspection

Visual inspection is the most traditional and easiest way of judging the quality of image data in cases where metrics like FID do not align with human perception. Visual inspection allows researchers to identify specific issues such as blurriness or distortions which are not captured by computational evaluation metrics. In our approach, we generated the same number of synthetic images using both the WGAN and DCGAN with the same model parameters trained over the same number of epochs. Figure 7 shows examples of the images from the original dataset compared with the images generated by our two GAN models. We can see that while the sample images produced by the DCGAN exhibit sharper clarity, those generated by the WGAN demonstrate greater diversity and closer resemblance to the original dataset when contrasted with the DCGAN-generated images.

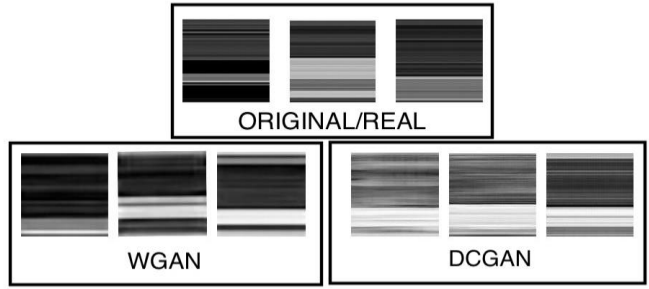

Figure 7. Examples of malware images generated: WGAN vs DCGAN

**5.1.2 Frechet Inception Distance**

As mentioned in section 3.2, The Frechet Inception Distance (FID) evaluates the quality of images generated by generative adversarial networks and how similar they are to the real ones. The lower the FID, the higher the quality of the images. We incorporated the FID∞ (a bias-free form of FID introduced in [34]) as part of both the DCGAN and WGAN generators. For the rest of the paper, FID refers to FID∞. We tasked both the DCGAN and WGAN generators, independently, to keep generating images until our machine ran out of memory. We generated a thousand images at every iteration, calculated the FID score for every image, and after removing anomalies, the calculated FID score is the average FID score of all the generated images for each one of the four datasets CICMalDroid 2020, Drebin, MalRadar, AndroZoo named Dataset 1, 2, 3 and 4 respectively. As seen in Figure 8-11, the FID scores of all WGAN-generated datasets reach lower values a little faster than those of DCGAN-generated datasets, showing that the WGAN-generated images are of slightly superior quality. We can all see that the WGAN model stops running at about 1000 epochs while the DCGAN model keeps running until about 1200 epochs, suggesting that the WGAN process takes a little more memory and effort than the DCGAN process for our particular design.

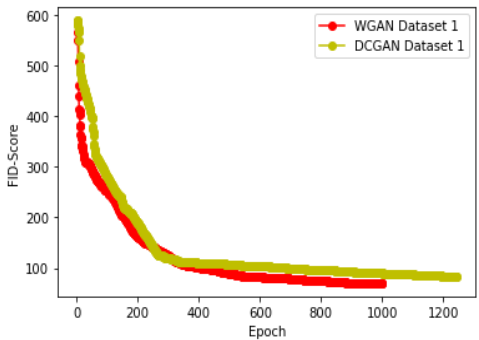 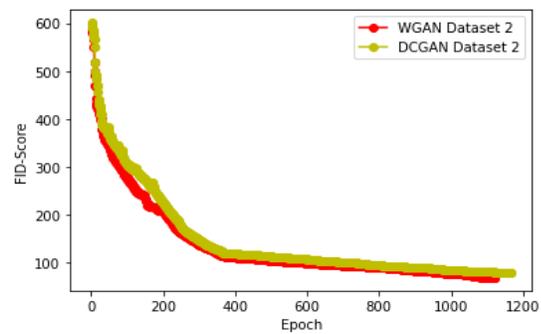

Figure 8. FID Score for Dataset 1           Figure 9. FID Score for Dataset 2

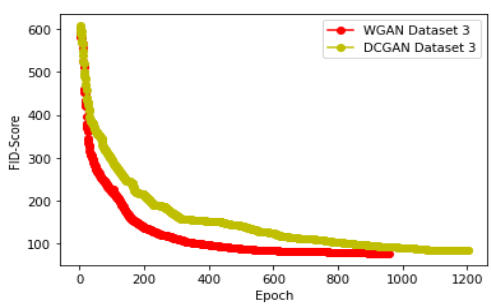 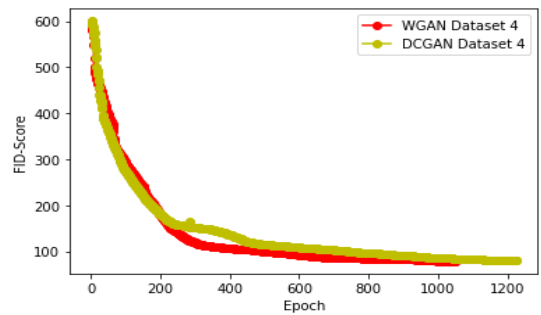

Figure 9. FID Score for Dataset 3           Figure 11. FID Score for Dataset 4

## 5.2 What Metrics Were Used to Evaluate the Performance of The Classification Model?

This study uses Confusion Matrix, also known as Error Matrix, to measure the performance of the classification model. Confusion Matrix can derive a variety of different indicators. The Confusion Matrix is composed of four values; TP, FP, FN, and TN as shown in Figure 10. Our datasets only have two categories; Malware and Benign. Therefore, in terms of judging whether a particular dataset of applications has malware, TP is the number of applications that the model predicted to be in the malware category and those applications indeed have malware. FP is the number of applications predicted by the model to be in the malware category but they do not actually have malware. FN is the number of malicious applications predicted by the model to be in the benign category but the applications actually have malware. TN is the number non-malicious applications predicted to be in the benign category by the model and the applications are actually benign. Other performance indicators such as accuracy, precision, recall, specificity, and the F1-score can be generated from the confusion matrix.

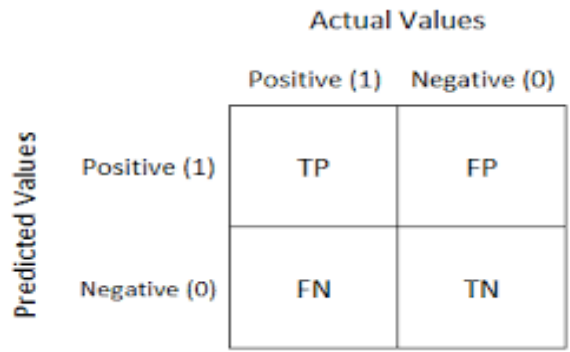

Figure 10. Confusion Matrix

## 5.3 How Does the Performance of The Classification Model Trained on Images Generated from Classes.dex Files Differ from That Trained on Images Generated from AndriodManifest.xml Files?

Table 1 shows the accuracy, precision, recall, F1 score, and specificity of the classification model trained on images of Classes.dex versus one trained on images of AndroidManifest.xml and tested on the never-seen dataset. For this particular task, we limited our comparison to only two datasets at image width 256x256. We discerned that achieving a comprehensive understanding could be accomplished in a cost-effective manner by focusing on these two datasets, rather than undertaking a comparison across all four datasets.

*Table 1. Performance of the classification model: images of Classes.dex vs images of Androidmanifest.xml*

| Dataset 1 Image source | Accuracy | Precision | Recall | F1-Score |
|---|---|---|---|---|
| Classes.dex | 0.955 | 00.970 | 0.952 | 0.961 |
| AndriodManifest.xml | 0.683 | 0.684 | 0.684 | 0.684 |
| **Dataset 4** | | | | |
| Classes.dex | 0.957 | 0.995 | 0.908 | 0.945 |
| AndriodManifest.xml | 0.671 | 0.626 | 0.646 | 0.636 |

From Table 1, we can see that the classification model trained on real images generated from the AndriodManifest.xml files performed badly, and combining them with those of Classes.dex files didn't help either. We noticed that images depicting malware and benign instances for Classes.dex files exhibit clear visual distinctions discernible to the naked eye, whereas the images associated with AndroidManifest.xml files lack such obvious differences, as mentioned in section 3.1. This suggests the features of the AndroidManifest.xml may not be suitable for malware detection. We conducted a more in-depth examination of this matter and identified the underlying reasons as follows:

- Limited Code Logic: The AndroidManifest.xml primarily provides static information about the app's structure and configuration. It does not contain the actual code logic or dynamic behaviors of the application.
- Dynamic Code Generation: Android apps can generate code dynamically during runtime, and this dynamic code may not be explicitly declared in the AndroidManifest.xml. Therefore, a complete understanding of the app's behavior requires analyzing the actual code, which is not present in the manifest file.
- Code Obfuscation: Obfuscation techniques can make it difficult to understand or duplicate the app's behavior solely by inspecting the AndroidManifest.xml.
- Permissions and Components Misuse: While the manifest file declares permissions and components, the actual usage of these permissions and components in the code may vary. Malicious apps, for example, may misuse declared permissions for nefarious purposes, which may not be evident from the manifest alone.
- Security Implications: Certain security-related aspects, such as encryption and data protection mechanisms, are typically implemented in the code rather than being explicitly specified in the manifest. Analyzing security features often requires inspecting the actual code.

**5.4. How Is the Performance of The Classification Model Trained on Real Data Alone?**

At this point in our study, our primary emphasis was directed towards the generation of image representations exclusively from the Classes.dex files. Solely utilizing authentic images for training the classification model, we observed that the highest F1 scores attained on the real yet previously unseen dataset were 0.961, 0.89, 0.930, and 0.945 for datasets 1, 2, 3, and 4, respectively, as illustrated in Figure 11.

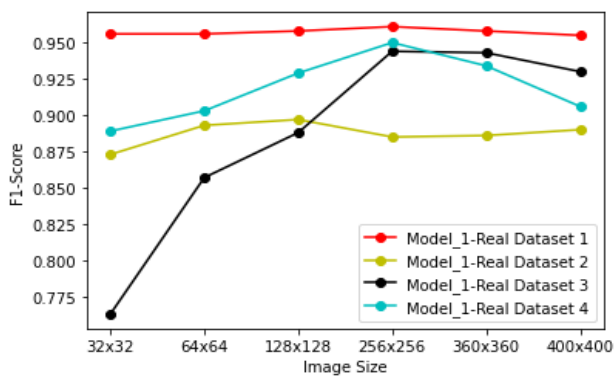

Figure 11. Model 1 F1-Score

**5.5 What Is the Difference in The Performance of The Classification Model Trained on WGAN Vs. DCGAN Generated Data Alone?**

Using only the images generated by the two GAN models to train the classification model, the highest F1 scores achieved by the classification model trained using the images generated by the WGAN to predict the real and previously unseen dataset were 0.88, 0.83, 0.76, and 0.80, (Dataset 1, 2, 3 and 4 respectively), and the highest F1 score achieved by the classification model trained on the images generated by the DCGAN were 0.84, 0.73, 0.70, and 0.77 (Datasets 1, 2, 3 and 4 respectively). The classification model trained on the WGAN-generated dataset performed better than that trained on DCGAN-generated for all datasets as shown in Figure 12-17 for datasets 1, 2, 3, and 4 respectively.

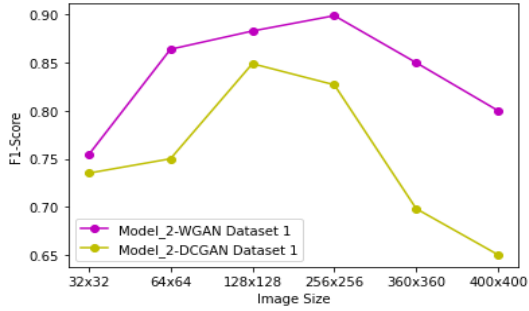

Figure 12. F1 Score of Classification Model Trained WGAN & DCGAN Dataset 1

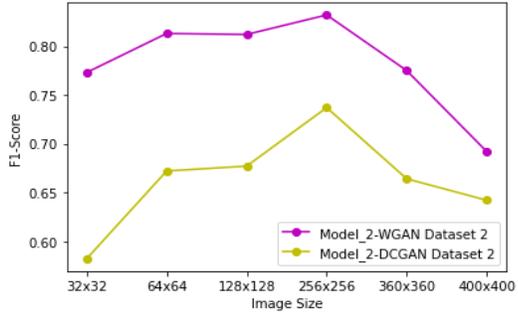

Figure 15. F1 Score of Classification Model Trained on on WGAN & DCGAN Dataset 2

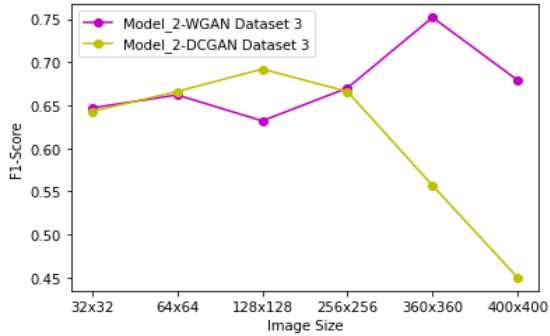

Figure 13. F1 Score of Classification Model Trained WGAN & DCGAN Dataset 3

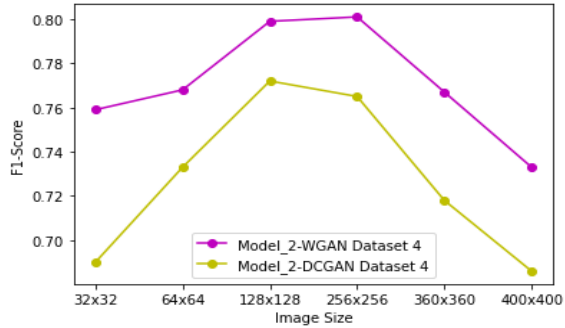

Figure 17. F1 Score of Classification Model Trained on on WGAN & DCGAN Dataset 4

## *5*.6. How Is the Performance of The Classification Model Trained on Real Data Combined with the GAN Generated Data?

At this point, we find ourselves exclusively relying upon the data generated by the WGAN for model training, owing to its demonstrated superiority in our specific context. Employing the amalgamated dataset, comprising both authentic and synthetically generated images produced by the WGAN, to train the classification model, the F1 scores attained on the previously unseen dataset were 0.975, 0.964, 0.946, and 0.972 for datasets 1, 2, 3, and 4, respectively, as shown in Figure 14.

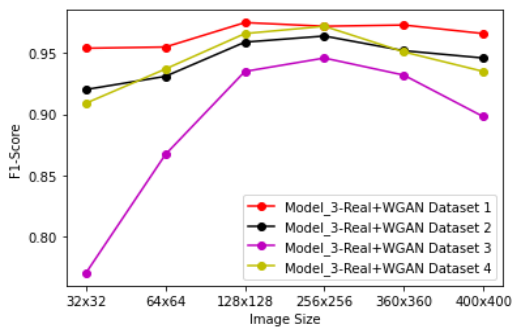

Figure 14. Model 3 F1-Score

## 5.7 What Is the Effect of Obsfucation On the Performance of The Classification Model?

As mentioned in section 3.1, obfuscation was achieved using the modular Python Android app obfuscator tool Obfuscapk. As long as at least 50% of obfuscated apps were used in the training and synthetic image generation processes, the performance of the classification model was only affected by negligible amounts.

## 5.8. Summary of Experimental Results

Table 2-7 show summaries of the classification model's performance for each image size respectively. The term 'Real data' refers to the image datasets generated from real malware samples, 'WGAN' refers to image datasets generated by the WGAN, and 'Real+WGAN' refers to combined datasets.

Table 2. Results at image size 32x32

| Image Source | Accuracy | Precision | Recall | F1-Score |
|---|---|---|---|---|
| **Real data** | | | | |
| *Dataset 1* | 0.949 | 0.954 | 0.957 | 0.956 |
| *Dataset 2* | 0.860 | 0.876 | 0.869 | 0.872 |
| *Dataset 3* | 0.753 | 0.738 | 0.790 | 0.763 |
| *Dataset 4* | 0.920 | 0.974 | 0.818 | 0.889 |
| **WGAN** | | | | |
| *Dataset 1* | 0.589 | 0.583 | 0.995 | 0.735 |
| *Dataset 2* | 0.775 | 0.872 | 0.694 | 0.773 |
| *Dataset 3* | 0.693 | 0.774 | 0.550 | 0.643 |
| *Dataset 4* | 0.840 | 0.938 | 0.638 | 0.759 |
| **Real+WGAN** | | | | |
| *Dataset 1* | 0.948 | 0.957 | 0.952 | 0.954 |
| *Dataset 2* | 0.914 | 0.951 | 0.891 | 0.920 |
| *Dataset 3* | 0.778 | 0.804 | 0.740 | 0.770 |
| *Dataset 4* | 0.934 | 0.994 | 0.837 | 0.909 |

Table 3. Results at image size 64x64

| Image Source | Accuracy | Precision | Recall | F1-Score |
|---|---|---|---|---|
| **Real data** | | | | |
| *Dataset 1* | 0.950 | 0.960 | 0.952 | 0.956 |
| *Dataset 2* | 0.882 | 0.886 | 0.900 | 0.893 |
| *Dataset 3* | 0.859 | 0.875 | 0.840 | 0.857 |
| *Dataset 4* | 0.930 | 0.985 | 0.835 | 0.903 |
| **WGAN** | | | | |
| *Dataset 1* | 0.750 | 0.878 | 0.655 | 0.750 |
| *Dataset 2* | 0.816 | 0.925 | 0.726 | 0.813 |
| *Dataset 3* | 0.708 | 0.783 | 0.580 | 0.666 |
| *Dataset 4* | 0.846 | 0.943 | 0.647 | 0.768 |
| **Real+WGAN** | | | | |
| *Dataset 1* | 0.949 | 0.970 | 0.940 | 0.955 |
| *Dataset 2* | 0.926 | 0.969 | 0.895 | 0.931 |
| *Dataset 3* | 0.869 | 0.885 | 0.850 | 0.867 |
| *Dataset 4* | 0.953 | 0.996 | 0.885 | 0.937 |

Table 3. Results at image size 128x128

| Image Source | Accuracy | Precision | Recall | F1-Score |
|---|---|---|---|---|
| **Real data** | | | | |
| *Dataset 1* | 0.952 | 0.953 | 0.963 | 0.958 |
| *Dataset 2* | 0.883 | 0.876 | 0.918 | 0.897 |
| *Dataset 3* | 0.889 | 0.897 | 0.880 | 0.888 |
| *Dataset 4* | 0.948 | 0.991 | 0.875 | 0.929 |
| **WGAN** | | | | |
| *Dataset 1* | 0.849 | 0.996 | 0.741 | 0.849 |
| *Dataset 2* | 0.816 | 0.928 | 0.722 | 0.812 |
| *Dataset 3* | 0.723 | 0.784 | 0.620 | 0.692 |
| *Dataset 4* | 0.863 | 0.953 | 0.687 | 0.799 |
| **Real+WGAN** | | | | |
| *Dataset 1* | 0.971 | 0.976 | 0.974 | 0.975 |
| *Dataset 2* | 0.956 | 0.983 | 0.936 | 0.959 |
| *Dataset 3* | 0.894 | 0.869 | 0.930 | 0.898 |
| *Dataset 4* | 0.974 | 0.997 | 0.937 | 0.966 |

Table 5. Results at images size 256x256

| Image Source | Accuracy | Precision | Recall | F1-Score |
|---|---|---|---|---|
| **Real data** | | | | |
| *Dataset 1* | 0.955 | 0.970 | 0.952 | 0.961 |
| *Dataset 2* | 0.873 | 0.884 | 0.886 | 0.885 |
| Dataset 3 | 0.944 | 0.949 | 0.940 | 0.944 |
| Dataset 4 | 0.957 | 0.995 | 0.908 | 0.945 |
| **WGAN** | | | | |
| Dataset 1 | 0.831 | 0.998 | 0.707 | 0.827 |
| Dataset 2 | 0.837 | 0.970 | 0.729 | 0.832 |
| Dataset 3 | 0.723 | 0.846 | 0.550 | 0.666 |
| Dataset 4 | 0.865 | 0.951 | 0.692 | 0.801 |
| **Real+WGAN** | | | | |
| Dataset 1 | 0.968 | 0.966 | 0.979 | 0.972 |
| Dataset 2 | 0.962 | 0.987 | 0.942 | 0.964 |
| Dataset 3 | 0.934 | 0.930 | 0.940 | 0.935 |
| Dataset 4 | 0.978 | 0.997 | 0.948 | 0.972 |

Table 4. Results at image size 360x360

| Image Source | Accuracy | Precision | Recall | F1-Score |
|---|---|---|---|---|
| **Real data** | | | | |
| *Dataset 1* | 0.950 | 0.978 | 0.933 | 0.955 |
| *Dataset 2* | 0.875 | 0.891 | 0.882 | 0.886 |
| *Dataset 3* | 0.944 | 0.968 | 0.920 | 0.943 |
| *Dataset 4* | 0.951 | 0.993 | 0.882 | 0.934 |
| **WGAN** | | | | |
| *Dataset 1* | 0.720 | 0.915 | 0.564 | 0.698 |
| *Dataset 2* | 0.785 | 0.915 | 0.672 | 0.775 |
| *Dataset 3* | 0.673 | 0.872 | 0.410 | 0.557 |
| *Dataset 4* | 0.845 | 0.943 | 0.646 | 0.767 |
| **Real+WGAN** | | | | |
| *Dataset 1* | 0.969 | 0.977 | 0.968 | 0.973 |
| *Dataset 2* | 0.949 | 0.974 | 0.931 | 0.952 |
| *Dataset 3* | 0.929 | 0.898 | 0.970 | 0.932 |
| *Dataset 4* | 0.963 | 0.995 | 0.911 | 0.951 |

Table 7. Results at image size 400x400

| Image Source | Accuracy | Precision | Recall | F1-Score |
|---|---|---|---|---|
| **Real data** | | | | |
| *Dataset 1* | 0.948 | 0.960 | 0.950 | 0.955 |
| *Dataset 2* | 0.879 | 0.893 | 0.886 | 0.890 |
| *Dataset 3* | 0.929 | 0.912 | 0.949 | 0.930 |
| *Dataset 4* | 0.931 | 0.981 | 0.842 | 0.906 |
| **WGAN** | | | | |
| *Dataset 1* | 0.903 | 0.999 | 0.831 | 0.907 |
| *Dataset 2* | 0.723 | 0.896 | 0.563 | 0.692 |
| *Dataset 3* | 0.608 | 0.761 | 0.320 | 0.450 |
| *Dataset 4* | 0.826 | 0.927 | 0.607 | 0.733 |
| **Real+WGAN** | | | | |
| *Dataset 1* | 0.962 | 0.975 | 0.958 | 0.966 |
| *Dataset 2* | 0.988 | 0.966 | 0.927 | 0.946 |
| *Dataset 3* | 0.944 | 0.923 | 0.970 | 0.946 |
| *Dataset 4* | 0.951 | 0.982 | 0.892 | 0.935 |

## 6. Conclusion

In this study, we have investigated the efficacy of using synthetic data to represent Android malware and reduce the amount of storage space needed. We've also compared differences in data quality between the WGAN and DCGAN generated data as well as the differences in performance of the CNN trained on real images with another trained on synthetic images generated by the GAN. Our data augmentation approach proved feasible, straightforward, and effective enough to be used in neural network training to detect real-world Android malware applications. Our results have shown that the model trained on WGAN-generated data is more effective than one trained on DCGAN-generated data and that combining synthetic data with real data is more effective than just using synthetic data or real data alone. Previous research on data augmentation for Android malware detection didn't use any mechanism to accurately evaluate the performance of their GAN models or the quality of the generated images. In contrast, in our experiments, image representations of malware are generated by two GAN models, and their quality is compared using the Frechet Inception Distance (FID) at every iteration. We've also compared our approach against recent studies in image representation and data augmentation for Android malware detection. As shown in Table 5, our approach utilizes more recent datasets and achieved a higher F1 score compared to Yi-Ming Chen et al [20] and Nadia Daoudi et al [15].

Table 5. Comparisons with recent research

|  | Our approach | Yi-Ming Chen et al | Nadia Daoudi et al |
|---|---|---|---|
| **Highest F1-Score** | 0.975 | 0.937 | 0.960 |
| **Datasets** | CICMalDroid 2020 | AMD | AndroZoo |
|  | Drebin | Drebin |  |
|  | MalRadar |  |  |
|  | AndroZoo |  |  |
| **Data Augmentation Technique** | DCGAN WGAN | DCGAN | None |
| **Static Analysis** | Yes | Yes | Yes |
| **Dynamic Analysis** | No | No | No |
| **Obfuscation** | Yes | No | Yes |
| **Method of GAN Performance Evaluation** | FID | None | Inapplicable |

## 6.1 Limitations and Threats to Validity

The following are the limitations and threats to the validity of this study:

1. Due to lack of time, storage space, and accessibility, we couldn't use the same amount of applications from the exact datasets as the recent studies mentioned above to compare. The limited size of our datasets might not accurately represent real-world malware, potentially introducing biases into the classification model.
2. Our image generation techniques couldn't generate images larger than 128x128 directly from applications fast enough due to a lack of computational power. We had to reshape them to 256x256, 360x360, and 400x400, which caused distortion. This could be the reason why the F1 score started dropping at images larger than 128x128 or 256x256.
3. Our study only focuses on static analysis. Static analysis tools may provide valuable insights into the app's structure, but they need to be complemented with dynamic analysis for a more thorough examination of an application's behavior and security posture. Wang Chao et al [43], showed that coupling static with dynamic analysis can significantly improve the accuracy of vulnerability mining of Android applications.
4. This research only considers the code in DEX files and permissions inside the manifest file. There's a possibility that applications can exhibit malware behavior inside other files within the application such as the META-INF and CERT.RSA files [44].

In the future, we would like to extend our study to include dynamic analysis as well as family classification.

## 7. Discussion

The future of image-based malware representation and Generative Adversarial Networks (GANs) in the realm of cybersecurity holds significant promise and potential advancements. Here are a few key considerations for the future: Researchers are likely to delve deeper into refining GANs for adversarial training, enabling models to better recognize and adapt to evolving malware obfuscation techniques. This could enhance the robustness of image-based malware representation, making detection models more resilient against sophisticated attacks. Continued advancements in deep learning and GANs may lead to improved feature extraction from image-based representations of malware. This could result in more nuanced and accurate models capable of identifying subtle patterns indicative of malicious activity. GANs are likely to play a pivotal role in generating synthetic datasets for training machine learning models. This could address challenges associated with limited real-world malware samples, and storage issues and contribute to the development of more comprehensive and diverse datasets. Image-based models could complement existing methods. This hybrid approach could leverage the strengths of both image-based and code-based analysis, providing a more holistic understanding of malware behavior. However, as the use of complex models like GANs becomes more widespread, there may be an increased focus on research into explainability and interpretability. Understanding how these models arrive at their decisions is crucial for building trust in their effectiveness.

**Acknowledgments**
We would like to thank the reviewers in advance for their guidance and comments.